# EXAMINING THE ROLE OF CONTEXT IN STATISTICAL LITERACY OUTCOMES USING AN ISOMORPHIC ASSESSMENT INSTRUMENT

SAYALI PHADKE
*Pennsylvania State University*
*sayalip@psu.edu*

MATTHEW BECKMAN
*Pennsylvania State University*
*beckman@psu.edu*

KARI LOCK MORGAN
*Pennsylvania State University*
*klm47@psu.edu*

**ABSTRACT**

*The central role of statistical literacy has been discussed extensively, emphasizing its importance as a learning outcome and in promoting a citizenry capable of interacting with the world in an informed and critical manner. Our work contributes to the growing literature on assessing and improving people's statistical literacy vis-à-vis contexts important in their professional and personal lives. We consider the measurement of contextualized statistics literacy - statistical literacy as applied to relevant contexts. We discuss the development of an isomorphic instrument modifying an existing assessment, design of a pilot study, and results which conclude that 1) the isomorphic assessment has comparable psychometric properties, and 2) test takers have lower statistical literacy scores on an assessment that incorporates relevant contexts.*

## 1. INTRODUCTION

The importance and role of statistical literacy has been discussed extensively in the statistics education literature (Ben-Zvi et al., 2018; Ben-Zvi & Garfield, 2004, 2008; Engel, 2017; Gal, 2002; Garfield et al., 2010; Gould, 2017; Rumsey, 2002; Schield, 1999; Utts, 2021; Wallman, 1993; Watson, 1998; Watson & Callingham, 2003; Weiland, 2017). Guiding documents which inform researchers and practitioners alike, such as the GAISE College Report (2016), Consortium for the Advancement of Undergraduate Statistics Education (CAUSE) Research Report (Pearl et al., 2012), International Handbook of Research in Statistics Education (Ben-Zvi et al., 2018), and GAISE PreK-12 Report (2020), highlight this importance vis-a-vis cognitive outcomes, curriculum, teaching practices, and assessments. The American Statistical Association discusses '(to) build a statistically literate society' as one of its objectives under the strategic goal of statistics education (ASA, n.d.). In parallel, the PARIS21 partnership (PARIS21, n.d.) among global organizations including United Nations, European Union, Organisation for Economic Co-operation and Development, International Monetary Fund, and the World Bank also considers statistical literacy to be a focus of its work. Even though definitions of statistical literacy vary in some aspects (Sharma, 2017), mainstream conceptualizations of statistical literacy agree that it comprises of a skillset which an individual would benefit from applying to contexts outside of a classroom, a skillset which would allow people to engage with contexts relevant to them from a data-driven point of view. Further, the literature also converges on a firm belief that statistical literacy plays a critical role in promoting a citizenry that is more capable of understanding the world around them and making evidence-based decisions in their private and public lives. Under this premise, our work asks the following question: Are students of statistics able to make sense of statistical insights encountered in their day-to-day lives, especially pertaining to relevant topics? Such an ability is considered to be an important marker of a statistically literate citizen (Wilks, 1951).

**1.1. ROLE OF CONTEXT AND TRANSFER**

Considerable amount of work has discussed the value of contexts and powerful ways of introducing contexts which are familiar to the students into the curriculum (DASL, n.d.; Gal, 2019; Garfield et al., 2012; Lee & Tran, 2015; Ratnawati et al., 2020). Concurrently, studies focusing on improving statistical literacy among students at various levels have also been conducted (Barbieri & Giacché, 2006; Carmichael, 2010; Ferligoj, 2015; Schield, 2004; Suhermi & Widjajanti, 2020; Watson, 2011). However, even though previous work has measured statistically literate behavior outside of a classroom setting (Budgett & Pfannkuch, 2007; Kaplan & Thorpe, 2010), there is limited work proposing research-based assessments of statistical literacy (Sabbag et al., 2018; Sanchez, 2007; Ziegler & Garfield, 2018). Assuming that applying statistical literacy skills to new contexts would involve a knowledge transfer (NASEM, 2000; Gal, 2003; Lovett & Greenhouse, 2000), we distinguish statistical literacy skills from the the ability to apply those skills to topics relevant in our lives, and define contextualized statistical literacy. Such a transfer, though it is central to the purpose of statistical literacy, is not encoded in the definition of statistical literacy. Contextualized statistical literacy is statistical literacy as it pertains to relevant contexts, where relevant contexts are conceptualized as ones that are societally relevant at a given time and people would have engaged with and thought about on their own. The key contribution of this work is in creating an instrument to measure contextualized statistical literacy using an existing research-based assessment of statistical literacy allowing us to examine respondents' statistical literacy skills when they are required to apply those in relevant contexts.

As underscored by Beckman (2015), the terms near and far transfer are relative and the distance of transfer implied in those terms is open to interpretation. Paas (1992) discusses distance vis-a-vis the similarity to problems encountered during instruction. Perkins & Salomon (1992) highlight that distance of transfer is an intuitive notion and discuss it as a matter of similarity and familiarity. Near transfer is across contexts students can be expected to be familiar with because they have encountered similar contexts before during instruction or practice. Whereas far transfer involves transfer across contexts which may not be similar, on the surface, to anything students have encountered before. According to Lovett & Greenhouse (2000), any application of statistical literacy skills can be considered to be a transfer problem. However, when considering contextualized statistical literacy, the question of distance of transfer is not straightforward. On one hand, encountering statistical constructs in new contexts increases the distance of transfer. On the other hand, though, irrespective of whether or not these relevant contexts have been introduced in the classroom before, since we conceptualize relevance as familiarity and engagement outside of the classroom, it can be considered to be nearer transfer for a respondent of an assessment of contextualized statistical literacy. When considering this transfer, we must also be also mindful of possible suspension of sense-making (Bonotto, 2002; Greer et al., 2007) whereby familiarity with the context maybe foregone in favor of focusing on the underlying statistical idea.

**1.2. ISOMORPHIC ASSESSMENT**

To measure the transfer of statistical literacy skills to relevant contexts, we created an isomorphic version of an existing research-based assessmet of statistical literacy. An isomorphic question or item is identical to a base item in structure (concept, phrasing, as well as distractors) and differs only in the context, continuing to measure the same underlying construct (Gick & Holyoak, 1980; Millar & Manoharan, 2021; Williamson et al., 2002). Isomorphic items can also be visualized as items with a common base template differing only in context (Beckman, 2015). Lehrer & Schauble (2007) and Fay et al. (2018) refer to these as structural isomorphs to highlight that this framework itself does not guarantee that respondents' cognitive processes in answering these tasks will be comparable. Isomorphic tasks have been studied extensively in the physics education literature (Barniol & Zavala, 2014; Kusairi et al., 2017, 2020; Lin & Singh, 2011; Luger & Bauer, 1978; Suganda et al., 2020). Some work has also been conducted in the computer science education domain (Millar & Manoharan, 2021; Parker et al., 2016). It is worth noting that there is limited work in the statistics education research literature which studies isomorphs. Most of the aforementioned studies deployed isomorphs to gauge learning and understand common misconceptions, and designed the study in such a way that each respondent solved all of the two or more isomorphic problems at different time points in a random order. Barniol & Zavala (2014) study was the only exception where each respondent was assigned to one of the two versions of the assessment. Previous work using isomorphic tasks finds that transfer across such tasks is difficult in most circumstances even if only incidental features are switched. There is some evidence, e.g. Kusairi et al. (2017), that more practice on the base topics improves performance as discussed by Lovett & Greenhouse (2000). Bassok & Holyoak (1989)'s findings are valuable given the objective of this work. Their work studied transfer across contexts which cross the disciplinary boundaries in which the construct is situated. They studied the effects of algebraic training on contexts within mathematics as well as in physics to find that training in mathematics facilitated transfer to physics but not the other way round. Even though every context in a statistics problem is external to the discipline itself, this work is important to consider because it provides some evidence of facilitating transfer where contextual information has been provided. This would indicate that familiarity with relevant contexts should improve student performance on statistical literacy tasks as compared to an isomorph based on potentially unfamiliar tasks barring any suspension of sense making (Bonotto, 2002; Greer et al., 2007).

These studies of transfer using isomorphic tasks deploy a variety of types of assessments. However, very few of them use research-based assessments. Parker et al. (2016) discuss the importance of developing assessment instruments which undergo rigorous process of collecting reliability and validity evidence, and for researchers to adopt these for further research. Cook & Hatala (2016), in outlining 'a practical approach to validation' of research-based assessments support the value and importance of this in step 4 – 'Identify candidate instruments and/or create/adapt a new instrument' - with a reminder to first look for previously developed instruments. We chose the Basic Literacy in Statistics (BLIS) assessment (Ziegler, 2014; Ziegler & Garfield, 2018) because of it's sole focus on statistical literacy and the extensive research conducted to gather reliability evidence and develop a validity argument for its intended use. We created an isomorphic version of

BLIS, i.e., M-BLIS hereafter, to answer the following research questions: (RQ1) Can an isomorphic instrument measure the same underlying construct as the original if all isomorphic items are dependent on relevant contexts? (RQ2) Do students perform comparably on both these assessments?

In Section **Error! Reference source not found.** we discuss the development of M-BLIS and design of the study implemented to gather reliability evidence and to develop a validity argument. Section 3 discusses results from the pilot data, concluding with a discussion of limitations and future research opportunities based on this work in Section 4.

## 2. METHODOLOGY

This section discusses the development of M-BLIS, design of the pilot study, and statistical methods used to analyze data from the study. Section 2.1 details the process of creating M-BLIS, including the choice of relevant contexts and parameters considered when developing isomorphic items. Additionally, it provides examples of modified tasks. We then outline (Section 2.2) the expert review process and the pilot study conducted at a large public research university in the eastern United States. Data from the pilot study are analyzed with the two research questions (RQs) outlined above. Finally, Section 2.3 describes the analytical methods.

### 2.1. ASSESSMENT MODIFICATION

Once BLIS was chosen as the instrument for measuring statistical literacy, it was modified to include isomorphic items. These isomorphic items were intended to depend on 'relevant contexts' - contexts which are societally relevant at the time and test-takers would have engaged with on their own outside of, and apart from, class. Various topic options such as climate change, immigration, race-related issues, and the COVID-19 pandemic were considered. The pandemic impacted the life of all individuals, presenting a unique opportunity for research in the form of a topic everyone could engage with and find relevant. We acknowledge the devastating effects of the pandemic and the psychological impact this may have on individual test-taker's performance. If a test-taker's performance is adversely affected by their emotional reaction to a given context (e.g., due to traumatic hardship or death of a loved one), we have limited ability to separate out the effect of the emotional reaction from their conceptual understanding. We must also consider the ethics of compelling a test-taker to look at statistics about a potentially sensitive topic such as the number of deaths due to the pandemic. Having said that, this issue is not unique to the COVID-19 pandemic. There is a broader question of whether and how to incorporate potentially sensitive topics on assessments while balancing the competing goals of encouraging students to look at relevant societal issues from a statistical lens on one hand, and the ethical and measurement-related issues arising as a result of the sensitive topic affecting test-takers differentially on the other hand. For the purpose of this work, we decided to proceed with the COVID-19 pandemic as the broader context with a clear intention of excluding any statistics pertaining to severe illness, loss of life or livelihood, and other serious health effects including but not limited to mental health. While mortality and hospitalization during the COVID-19 pandemic were almost certainly engaging contexts, the strong negative association for students who may have endured such trauma or loss would likely evoke an emotional response. Such contexts were avoided mainly as a matter of compassion. Additionally, though, such contexts could have prevented the respondent from completing the assessment to the best of their ability. Two competing requirements of improving engagement and reducing emotional impact were balanced during the process of creating the isomorphic assessment. Contexts such as sleep length among college students, dental care choices, flight cancellations, restaurant visit frequency, and air pollution, all during the COVID-19 pandemic, were incorporated.

Each item went through extensive considerations. Before looking for data and reports from which contexts and statistics were sourced, following were noted for each item on BLIS: type(s) of variable(s), parameter(s) of interest, type of sample, type of study (observational versus experimental), and whether creation of the item required access to raw data or summary statistics or neither. Additionally, keywords were considered in an attempt to find a context which could match the original item as closely as possible. Table 1 demonstrates an item for which an alternative study with a highly comparable context was found. The original item was based on a real but not widely relevant context. This item may be considered one of the purest forms of an isomorph in M-BLIS. The underlined text highlights common words across the two versions, excluding singulars or plurals.

*Table 1. Real from a real study to Real from a relevant study*

| Original item stem | Modified item stem |
|---|---|
| Dogs have a very strong sense of smell and have been trained to sniff various objects to pick up different scents. An experiment was conducted with a dog in Japan who was trained to smell bowel cancer in stool samples. In a test, the dog was presented with five stool samples; one from a cancer patient and four from healthy people. The dog indicated which stool sample was from the cancer patient. This was repeated a total of 38 times. Out of the 38 tests, the dog correctly identified the cancer sample 37 times. A hypothesis test was conducted to see if this result could have happened by chance alone. The alternative hypothesis is that the dog correctly identifies cancer more than one fifth of the time. The p-value is less than .001. Assuming it was a well-designed study, use a significance level of .05 to make a decision. | Dogs have a very strong sense of smell and have been trained to sniff various objects to pick up different scents. A pilot experiment was conducted with dogs in Germany who were trained to smell COVID-19 in saliva samples. In the test, one dog was presented with 115 saliva samples; 21 from COVID-19 patients and 94 from healthy people. The dog indicated which saliva samples were from the COVID-19 patients. Out of the 21 COVID-19 positive samples, the dog correctly identified 20 of them. A hypothesis test was conducted to see if this result could have happened by chance alone. The alternative hypothesis is that the dog correctly identifies COVID-19 more than half the times. The p-value is less than .001. Assuming it was a well-designed study, use a significance level of .05 to make a decision. Source: Research article. |

Each original item was also categorized based on the scale discussed in the GAISE College Report (2016): Naked data, Realistic data, Real data, Real data from a real study. This is reflected in the table caption. This categorization served a dual purpose. The first and broader purpose of considering this categorization was to analyze whether any observable effect is associated with with the degree of change from the original data category to the modified category (real data from a relevant study). Secondly, it allowed us to understand which isomorphs needed to go beyond simply replacing context-specific words. For example, the item in Table 2 was based on naked data in the original BLIS. However, given the purpose of this work, the change had to go beyond a simple isomorph.

*Table 2. Naked to Relevant*

| Original item | Modified item |
|---|---|
| The distribution for a population of measurements is presented below. | For scientific credibility, journal articles are reviewed by other scientists before |

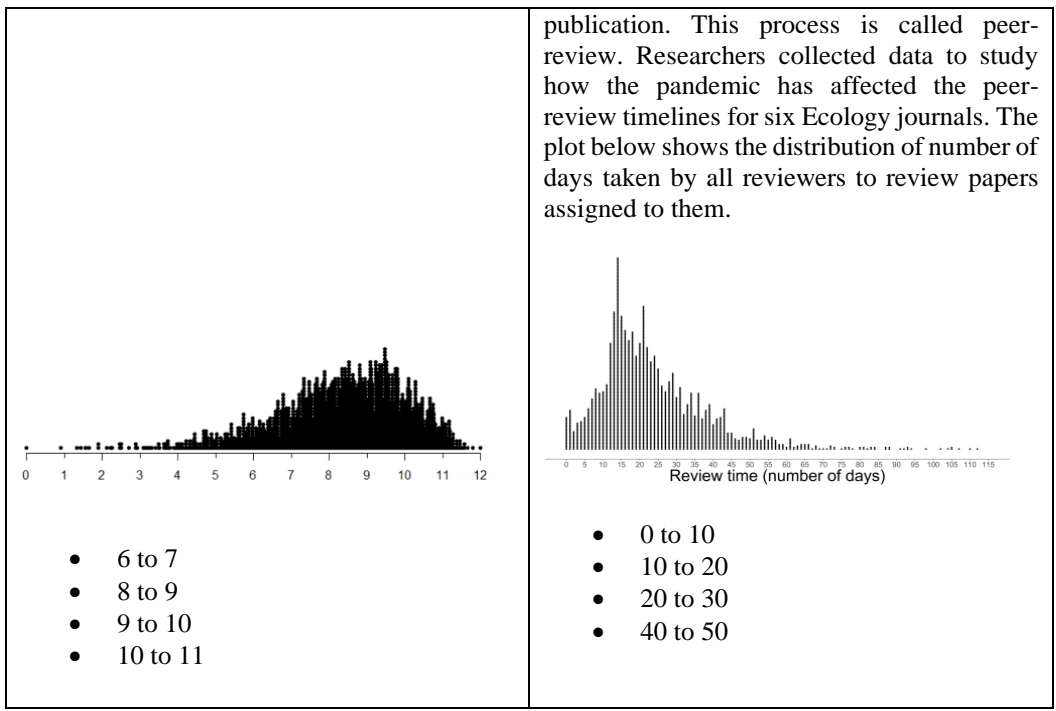

publication. This process is called peer-review. Researchers collected data to study how the pandemic has affected the peer-review timelines for six Ecology journals. The plot below shows the distribution of number of days taken by all reviewers to review papers assigned to them.

- 6 to 7
- 8 to 9
- 9 to 10
- 10 to 11

- 0 to 10
- 10 to 20
- 20 to 30
- 40 to 50

In contrast, Table 3 demonstrates an item which was based on real data from a real study, leading to an isomorph that retained the structure of the original item very closely.

*Table 3. Real from a real study to Relevant from a real study*

| Original item stem | Modified item stem |
|---|---|
| The Pew Research Center surveyed a nationally representative group of 1,002 American adults in 2013. Of these adults, 21% have had an email or social networking account compromised. Identify the population about which the Pew Research Center can make inferences from the survey results and the sample from that population. | The Pew Research Center surveyed a nationally representative group of 12,648 U.S. adults in November 2020. Of these adults, 62% said they would be uncomfortable being among the first to get the vaccine for COVID-19. Identify the population about which the Pew Research Center can make inferences from the survey results and the sample from that population. Source: Pew Research Center. |

For each modified item, the source link was provided at the end of the prompt. It was added on a separate line with the word "Source" followed by a very short key phrase identifying the source with a hyperlink. This was intended to underscore the authenticity and credibility of the contexts presented in the item without distracting the test-taker from the key task. During a think-aloud conducted prior to field testing, a respondent explicitly stated that this added legitimacy to the questions in the student's mind.

The stringent requirement to retain the structural integrity of item phrasing and the statistical idea in the isomorph was loosened for two items. The modified context for the item in Table 4 was powerful enough to compel such a concession.

*Table 4. Context changed considerably*

| Original item stem | Modified item stem |
|---|---|

| According to the National Cancer Institute, the probability of a man in the United States developing prostate cancer at some point during his lifetime is .15. What does the statistic, .15, mean in the context of this report from the National Cancer Institute? | Consider an individual fitting the following description.<br>• 20-year-old female,<br>• lives alone near a university campus,<br>• is exposed to an average of 10 people each week,<br>• has no underlying medical complications,<br>• is asymptomatic and unvaccinated, and<br>• follows CDC's guidance.<br>According to the "19andMe" tool developed by Mathematica, her probability of catching COVID-19 through community transmission in a week is .0024, as of March 30, 2021. What does the statistic, .0024, mean in the context of this calculation from Mathematica? Source: Online calculator. |
|---|---|

Finally, the item in Table 5 was a subject of lengthy discussions, some of which included the expert reviewers. The implicit assumption of a coin being unbiased and our intuition about 50% of them landing on heads benefitted the original item. However, upon deliberation, it was agreed that it is extremely hard to find other phenomena which have an unconditional 0.5 probability of occurrence which is understood intuitively, and therefore the substantial change in wording was included.

*Table 5. Implicit assumption changed*

| **Original item stem** | **Modified item stem** |
|---|---|
| Two students are flipping coins and recording whether or not the coin landed heads up. One student flips a coin 50 times and the other student flips a coin 100 times. Which student is more likely to get 48% to 52% of their coin flips heads up? | Penn State University administrators surveyed all undergraduate students to capture feedback from the entire student body on several issues. As a result, they learned that 86% of all students planned to return in fall 2020. Despite knowing the proportion for all Penn State students as a whole, several instructors surveyed their own classes in order to be sensitive to the views of their students. One instructor had a class with 50 students and another instructor had a class with 100 students. Assuming both classes were representative of the entire student body at Penn State, which instructor was more likely to find that 84% to 88% of their students would plan to return in fall 2020? Source: Adapted from Penn State News. |

In addition to the factors discussed above, we also considered peculiarities such as the distance of the sample statistic from the parameter, scale of the p-value, whether a small sample was required, and overall length (in characters) of the names of variables or context description. The original assessment was unchanged from the version provided in Ziegler (2014). For items that involved a visualization, plots were created using the *ggplot2* package (Hadley Wickham, 2016) in *R* (R Core Team, 2021). Even though some of the original visualizations were created using a the *plotrix* package (J, 2006), same aesthetics and scales were maintained in the modified visualizations.

## 2.2. STUDY DESIGN

To investigate whether M-BLIS continues to measure the same underlying constructs and whether students perform comparably on both assessments, we gathered data to generate reliability evidence and develop a validity argument using methods recommended in the *Standards* (AERA, 2014). These included expert reviews, think aloud interviews, and a pilot study, though they were adapted to be suitable for the development of an isomorphic assessment instead of a new instrument.

Three expert reviewers looked at the modified instrument with the prompt, "Please consider each modified item vis-a-vis the original item and comment on whether they are comparable in measuring the underlying learning outcome." The instrument was updated based on expert feedback for a pilot study. In the pilot instrument, six out of the 37 total items were retained as anchors. This allowed us to equate scores under the internal-anchor design discussed in Livingston (2004).

This updated instrument was piloted in a study conducted at a large public research university in the eastern United States. The pilot was designed to address the two questions described above. (RQ1) How does the functioning of the isomorphic items compare to the functioning of the original assessment? (RQ2) Is there evidence to suggest that test-takers respond to the underlying statistical question differentially if the item is based on a relevant context? To facilitate this investigation, we built two levels of comparisons. First, consenting test-takers were randomly assigned to take either BLIS or MBLIS. This gave us a baseline on the original assessment within the target population, faciliating a comparison of results across the results from Ziegler (2014)'s field test and our pilot study. This allowed us to answer the main research question - is MBLIS measuring the same constructs as BLIS? To add another layer of comparison, five randomly chosen items from the original assessment were retained as anchors. One of the randomly selected item numbers was a part of a testlet leading to six anchor items. Resultantly, M-BLIS featured 6 original and 31 modified items. Alternative criteria such as model-based difficulty ranking and topic were considered for the determination of the anchor items. However, a random selection was decided to be the best choice at the end.

The original instrument was developed specifically for an undergraduate introductory statistics student learning under the simulation-based inference curriculum. Although learning under a different framework (Lock5 curriculum instead of CATALST), the undergraduate introductory statistics course at the aforemention university matched this description, providing in its students an easy choice of group to use as pilot subjects. The original instrument was studied as a mid-semester and end-of-semester evaluation. Therefore, the pilot study deployed it as a post test in the aforementioned class.

Finally, an extensive survey was attached at the end of the assessment to learn student demographics, their interest in and engagement with certain relevant topics such as diversity questions, immigration, politics and governance, and their experience of interacting with items pertaining to the pandemic (given only to M-BLIS respondents). The last subset included questions such as whether responding to items regarding the pandemic was discomforting to them and whether their ability to consider the statistical question was affected by the contexts.

The data collected from this pilot will be used to further validate M-BLIS. Learning about differential performance across the two assessments within similar subgroups of students will provide us with useful information on the performance of the modified instrument. Further, item-by-item comparison across the instruments will allow us to look more closely at the items which may perform differently. These findings will be most instrumental in us developing the validity and reliability argument for M-BLIS.

## 2.3. STATISTICAL METHODOLOGY

Data from the pilot study were analyzed with two key goals in mind. First, are the two assessments (BLIS and M-BLIS) comparable in measuring underlying constructs? Second, did the test-takers perform comparably on both the assessments? Though both sets of analyses were conducted using the same student performance data, separating out these two goals helps us discuss results accordingly.

To compare the two assessments themselves, we evaluated measures of reliability and measures which can contribute towards a validity argument for the use of the instrument. All the measures in this set are replicated based on the analyses in Ziegler (2014). As a measure of reliability, we consider the coefficient of alpha to compare internal consistency among items. In order to check the assumptions of Item Response Theory (IRT) models which contribute towards the validity argument, we conduct principal component analysis to check two assumptions of the IRT models - unidimensionality and local independence. Scree plots based on PCA allow us to comment on that. Single-factor confirmatory factor analysis allow us to further comment on unidimensionality. Finally, we also fit Rasch, 2PL, and PC models and look at item information curves from the best of these models to learn whether item difficulties are comparable across the two assessments. Test information functions and standard errors of measurement are also considered for each instrument. On the validity side, we look at item parameters for partial credit (PC) model with 32 items and 4 testlets, and item characteristic curves for each item or testlet. It is important to note that even though we may ocassionally compare our results to the original study conducted in form of field test in Ziegler (2014), a separate set of analyses where the pilot study is considered a replication of the original study will be discussed in forthcoming work. For the purpose of this discussion, we limit ourselves to analyses which consider whether the two instruments, as suggested by data from our pilot, function comparably with each other in terms of the reliability measures and pieces of the validity argument.

To compare student performance more directly, we analyzed data under the classical test theory (CTT) framework. Though we acknowledge the advantages of using the IRT framework when analyzing assessment data, CTT was preferred due to the ease with which the relationship between test-taker covariates and performance can be interpreted in the models. We fit multiple linear regression models - with total scores as the response variable - to understand the differential performance across two assessments. Type of assessment randomly assigned to a student was the key explanatory variable of interest. We also included student demographics such as their gender, whether they are an international student or not, and highest education of a parent/guardian, as well as responses to pertinent survey questions. These analyses were conducted using all responses, as well as the complete-only responses. In this paper, we present the latter.

## 3. RESULTS

## 3.1. COMPARING ASSESSENT INSTRUMENTS

In this section, we address the first research question (RQ1): do the two instruments, BLIS and M-BLIS, perform comparably?

***Summary of Assessment Performance*** First we summarize item-by-item performance to highlight key differences. Table 6: Difference in proportion of respondents correctly answering each itemTable 6 contains percentage of correct responses per item. The *Difference* column is calculated as $correct_{BLIS} - correct_{M-BLIS}$. Table 19 in the Appendix (Section 6.2) tabulates selected-responses i.e. percentages of respondents who chose each distractor. Shaded blue rows indicate anchor items which are critical in comparing the two groups of respondents at baseline. Shaded orange rows highlight items with difference values around or higher than an arbitrary cutoff of 10%.

*Table 6: Difference in proportion of respondents correctly answering each item*

| Item | BLIS | M-BLIS | Difference | BLIS context - GAISE |
|---|---|---|---|---|
| 1 | 74.6 | 73.2 | 1.4 | Real from real study |
| 2 | 44.0 | 50.7 | -6.7 | Realistic |
| 3 | 53.4 | 52.5 | 0.9 | Real |
| 4 | 83.5 | 86.2 | -2.7 | Real |
| 5 | 81.3 | 84.7 | -3.4 | Realistic |
| 6 | 73.5 | 70.7 | 2.8 | Realistic |
| 7 | 35.6 | 41.1 | -5.5 | Real from real study |
| 8 | 29.5 | 32.8 | -3.3 | Realistic |
| 9 | 65.4 | 34.0 | 31.4 | Realistic |
| 10 | 56.3 | 39.2 | 17.1 | Realistic |
| 11 | 42.0 | 37.1 | 4.9 | Real |
| 12 | 58.3 | 48.8 | 9.5 | Real from real study |
| 13* | 37.6 | 37.6 | 0.0 | Real from real study |
| 14 | 42.8 | 24.6 | 18.2 | Naked |
| 15 | 63.8 | 48.5 | 15.3 | Realistic |
| 16* | 24.6 | 27.8 | -3.2 | Realistic |
| 17* | 46.1 | 46.8 | -0.7 | Realistic |
| 18 | 45.9 | 45.4 | 0.5 | Real from real study |
| 19 | 40.8 | 38.4 | 2.4 | Real from real study |
| 20 | 37.9 | 34.3 | 3.6 | Real from real study |
| 21 | 16.5 | 16.3 | 0.2 | Real from real study |
| 22 | 58.5 | 61.0 | -2.5 | Realistic |
| 23* | 43.4 | 43.9 | -0.5 | Real from real study |
| 24* | 57.2 | 60.0 | -2.8 | Real from real study |
| 25 | 55.5 | 61.6 | -6.1 | Real from real study |
| 26 | 42.2 | 42.0 | 0.2 | Real from real study |
| 27 | 38.6 | 45.0 | -6.4 | Realistic |
| 28 | 52.7 | 60.2 | -7.5 | Real from real study |
| 29 | 52.0 | 48.9 | 3.1 | Real from real study |
| 30 | 48.3 | 45.5 | 2.8 | Real from real study |
| 31 | 86.4 | 83.9 | 2.5 | Realistic |
| 32* | 48.0 | 43.6 | 4.4 | Real from real study |
| 33 | 64.4 | 62.0 | 2.4 | Real from real study |
| 34 | 70.4 | 65.2 | 5.2 | Real |
| 35 | 23.4 | 21.6 | 1.8 | Real from real study |
| 36 | 79.0 | 68.6 | 10.4 | Real |
| 37 | 57.8 | 54.3 | 3.5 | Real from real study |

**Anchor items:** All anchor items (13, 16, 17, 23, 24, 32) had a difference of less than 5% in the proportion of respondents who chose the correct answer. On four of the five remaining items, two of which formed a testlet (*items 23-24*), the M-BLIS group had a higher percentage of respondents choosing the correct answer. On *item 32*, students in the BLIS group performed better. However, authors must note here that there was an inconsistency in the presentation of *item 16* across the two versions. Aside from this *item 16*, the distribution of selected responses was comparable across the two assessments, providing evidence of the two randomized groups being comparable at baseline.

**Isomorphic items:** Excluding the anchor items, 22 out of the remaining 31 items witnessed better performance on BLIS. Further, the items with the highest absolute difference were ones where the BLIS saw better performance. The items where students performed better on M-BLIS are unevenly distributed in terms of topic coverage. The latter half of the assessment was based on inferential statistics and it appears as though BLIS was easier for those topics. This gives us an early indication that items with relevant context may have been more difficult to answer. Having said that, about half the differences were less than 5%, suggesting that the instruments may be more comparable than suggested by the extreme difference values.

Six items had absolute differences close to or higher than 10%. Respondents performed better on BLIS on all of these items. Five out of these six items pertained to graphs and descriptive statistics, and the sixth was based on regression and correlation. The difficulty levels of these items, according to Ziegler (2014)'s analysis, were well-distributed. Two of these six items were discussed in Section 2.1. The BLIS item which was naked (Table 2) and the item where length was changed considerably to accommodate a pertinent relevant context (Table 4) were both in this group. Further, the item which changed from naked to relevant saw an incorrect option being chosen more frequently than the correct one, even though this observation must be treated with care since the direction of the skewness was reversed.

Item 10 warrants a closer look not only because of the high difference in proportion, but also because of the contexts. The BLIS context for this item was 'number of hours of sleep for college students,' whereas the M-BLIS context was 'an index capturing the strictness of lockdown policies implemented by various countries' governments in response to the COVID-19 pandemic.' This is also true, with a higher difference value, for item 9. The BLIS context was 'self-reported confidence about success of students in an introductory statistics class,' and the M-BLIS context was 'rating from Vietnamese citizens indicating how overwhelming the official news regarding the pandemic has been for them.' In contrast, though, four other items (5, 6, 8, and 35) also featured contexts specific to college students on BLIS and the differences are much smaller for those items.

Finally, we looked at two specific items of interest. Of the five testlets included in this instrument, only one shared a learning outcome - items 29 and 30. Item 29 required a respondent to choose the correct null hypothesis and 30 required them to select the alternative hypothesis for the same stem. Table 7 tabulates performance on these two items.

*Table 7. Performance on testlet items*

|  | BLIS | | M-BLIS | |
| --- | --- | --- | --- | --- |
|  | **Correct – Q30** | **Incorrect – Q30** | **Correct – Q30** | **Incorrect – Q30** |
| **Correct – Q29** | 40.9 | 11.1 | 35.1 | 13.8 |
| **Incorrect – Q29** | 7.4 | 40.6 | 10.4 | 40.7 |

In the original study, the author decided to treat items 29 and 30 as a testlet because majority of the students either got both questions right or both of them wrong. Since this observation holds true for M-BLIS as well, we treat our data as a 36-item-scale (with one testlet) for the remaining analyses.

***Reliability evidence and evidence for validity argument*** The expert reviews contribute to the evidence used to develop a validity argument for the intended use of M-BLIS. A large proportion of expert comments entailed a change in the structure of the original instrument itself. These comments ranged from minor changes in wording to discussions regarding whether the item is measuring the construct it claims to. However, any comment which would have led to a change in the original item was marked as out of scope for the purpose of this work. Of the 31 items (six anchors) under consideration, 12 were unchanged and 16 received minor wording and presentation changes. The language and presentation of one item and all its distractors was updated significantly. Two items were topics of lengthy discussions. They were almost completely rewritten based on expert reviews. One of these two items is presented in Table 5 and discussed subsequently. The other isomorph that underwent a significant change based on expert reviews shared a key characteristic with the item in Table 5. The context in that item also had a binary outcome which could be intuitively assumed to be equiprobable. In both these cases, the original context of our choice was retained while rephrasing the item stem. The resulting instrument based on these changes was deployed in the pilot study discussed in Section 2.2.

The coefficient alpha indicates internal consistency among test items. In Table 8, *alpha_36* is the raw coefficient alpha with the single testlet, *predicted_32* is the predicted coefficient alpha for the 32-item instrument using the Spearman-Brown formula, and *alpha_32* is the actual coefficient alpha for scores with the other 4 testlets. Scores for the other four testlet (excluding items 29-30 discussed in Table 7) include partial scoring. The small differences between the 36 and 32 item scales, as well as across the predicted and calculated values for the 32-item scale, indicate that local independence is not a concern among testlet items. These metrics were chosen in order to compare our results with those discussed in Ziegler (2014).

*Table 8. Raw and predicted coefficient alpha values*

|  | BLIS | M-BLIS |
|---|---|---|
| alpha_36 | 0.78 | 0.77 |
| predicted_32 | 0.76 | 0.75 |
| alpha_32 | 0.78 | 0.76 |

IRT analyses conducted to develop a validity argument make assumptions of unidimensionality in the assessment scale and local independence among items. Results of principle Component Analysis (PCA) are summarized as scree plots in Figure 1. The left panel displays scree plots for the 36-item scale and the right panel displays equivalent plots for the 32-item scale with 4 testlets. The two plots within each panel are very similar, effectively hiding the grey dots for BLIS. However, this is encouraging evidence indicating that the two assessment versions are performing comparably.

The eigenvalues level-off after the first factor providing support to the hypothesis that the assessment instruments both consist of a single factor. We do not observe any clear differences between the 36 and 32 item assessments. All the scree plots show evidence of unidimensionality in the instruments. Acceptability of the local independence assumption was checked using single-factor confirmatory analyses. Results indicated that including testlet scores is acceptable to meet the local independence assumption.

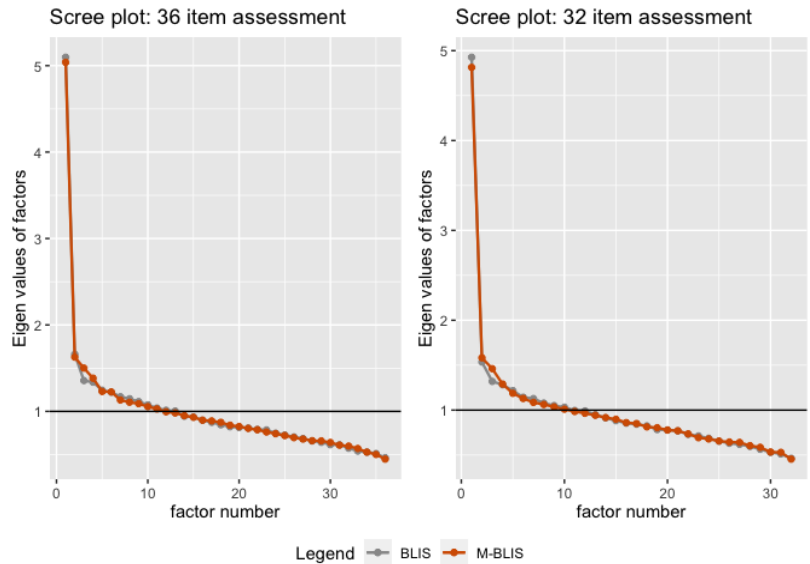

*Figure 1: Scree plots of eigenvalues from PCA*

Three IRT models were fit to these data - Rasch model, a 2 parameter logistic (2PL) model, and a partial credit (PC) model. The Rasch and 2PL models were based on 36-item scale, whereas the PC model was based on the 32-item scale incorporating partial grading on the four testlets. Table 9 summarizes three model fit measures i.e. the Akaike Information Criterion (AIC), Bayesian Information Criterion (BIC), and the Log Likelihood (LL). Based on these indices, the PC model seems to perform the best on both assessments, though the differences in values are fairly small. Therefore, the PC model will be used for the remainder of the analyses.

*Table 9. Model summaries for IRT models*

|  | BLIS | | | M-BLIS | | |
|---|---|---|---|---|---|---|
|  | Rasch | 2PL | PC | Rasch | 2PL | PC |
| AIC | 27771.08 | 27171.91 | 26358.63 | 26902.39 | 26246.05 | 25607.35 |
| BIC | 27936.04 | 27492.91 | 26523.58 | 27065.99 | 26564.41 | 25770.95 |
| LL | -13848.54 | -13513.95 | -13142.31 | -13414.20 | -13051.03 | -12766.67 |

Figure 4 and Figure 5 in the Appendix (Section 6.3) display item information curves. These curves indicate that the instruments contain items which give us information across the ability levels, thereby differentiating test-takers across all ability levels. When comparing the two assessments we notice that there are a few more items on the modified instrument measuring students at higher ability levels that those on the original scale. The test information function and standard error (SE) of measurement curves in Figure 6 and Figure 7 (Section 6.3) support this observation. Overall test information for M-BLIS is highest at a slightly higher ability level than the it is for the original instrument. For BLIS, the SE is slightly lower at lower ability levels, giving slightly more information at lower abilities. For M-BLIS, SE is comparable at the extremes, indicating that it is giving equally little information at the highest or lowest abilities.

Finally, difficulty rankings based on the PC model and item characteristic curves are considered for validity evidence. Table 20 in the Appendix (Section 6.3) displays difficulty estimates based on the PC model. They range from -2.02 to 1.78 for BLIS and from -1.98 to 1.78 for M-BLIS. Even though both assessments display comparable ranges of difficulty spread evenly on either both side of zero, the distribution of difficulties is slightly uneven on either side of zero. Half of the 32 items/testlets on the original instrument have difficulty estimates lower than zero. This division is 14 under zero and 18 above for the modified instrument. In line with earlier results, the five items which are most distant in terms of difficulty rankings when BLIS and M-BLIS are compared, are the same items which had higher than 10% difference in Table 6. Further, the two items with highest difficulty (consistent across the two instruments) are the two items for which respondents chose an incorrect option most frequently according to Table 19 in the Appendix (Section 6.2). These findings are also supported in the item characteristic curves seen in Figure 8 and Figure 9 also in the Appendix (Section 6.3). However, these items have negative correlations with the total score without accounting for the given item. This reverse discrimination indicates a possible flaw in the item and warrants further investigation.

### 3.2. COMPARING STUDENT PERFORMANCE

(RQ2): did respondents perform comparably on the two assessment instruments BLIS and M-BLIS? In this section, we address the second research question by exploring the relationship between assessment performance and assessment type. We also incorporate various test-taker attributes to further understand this relationship. Figure 2 shows a distribution of total score by assessment. Assessments are scored as one point per correctly answered question with a highest possible total of 37 points. The grey bars represent scores on the M-BLIS and yellow bars represent scores on BLIS.

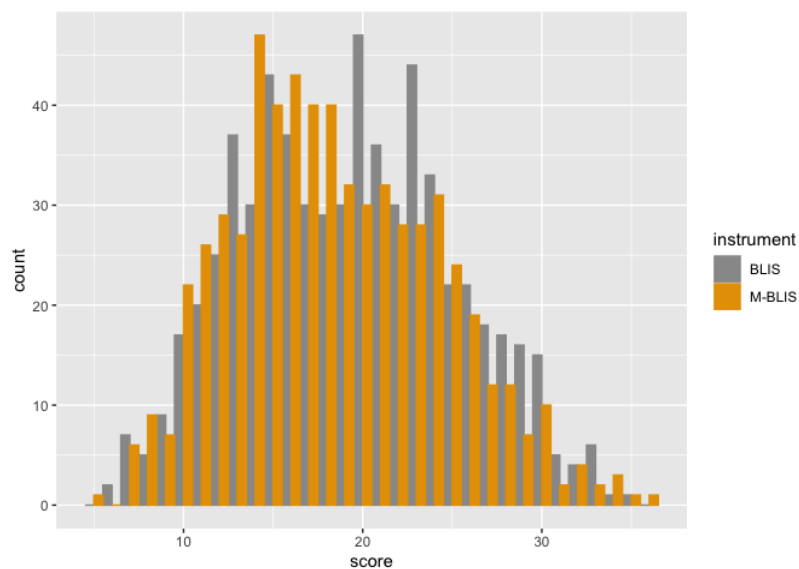

*Figure 2: Comparison of total score (out of 37)*

More students scored higher on BLIS. Figure 3 in the Appendix (Section 6.2) plots the two histograms in separate panels, highlighting the slight right skew in the scores on M-BLIS. Overall scores were comparable on both assessments, as suggested by the numerical summaries in Table 10. However, a two-sided t-test for the difference in mean scores led to a p-value of 0.005 indicating a significantly lower score on M-BLIS.

*Table 10: Summary statistics of total scores*

| Instrument | n | Mean | Median | SD | IQR |
|---|---|---|---|---|---|
| **BLIS** | 638 | 19.31 | 19 | 5.98 | 8.75 |
| **M-BLIS** | 615 | 18.38 | 18 | 5.78 | 9 |

Before looking at results from statistical models which consider the difference between BLIS and M-BLIS scores accounting for covariates, univariate summaries of important variables were considered to identify any categories with a small *n* (Table 13 - Table 18 in Appendix A (Section 6.1)). For all tables, non-responses were removed. There were less than (1%) missing values in any of the variables. Some categories of the gender variable and the group of students who expected to get an F in the class were the only groups with less than (1%) frequency. Bivariate frequencies of all these variables crossed with assessment type were also considered. They ascertained that the these variables have comparable distributions across the BLIS and M-BLIS groups.

Four survey questions were considered important for the statistical models in an effort explore whether they were related to assessment performance. A set of survey questions explicitly asked the respondent about their interaction with COVID-related contexts as well as whether the contexts interacted with their assessment-taking experience. Before including the survey responses into an inferential model, we summarize them. Table 11 tabulates percentage responses to three statements "I have actively looked for information on this topic (COVID-19) in the last 6 months." (**Engagement**), "I would like to gain data-driven insights into this topic (COVID-19)." (**Statistical interest**), "I think this topic (COVID-19) is relevant to our lives" (**Relevance**), each with "Yes", "Maybe", or "No" alternatives, based on the assessment type.

*Table 11. Survey questions regarding COVID-19 pandemic*

| | Engagement | | | Statistical interest | | | Relevance | | |
|---|---|---|---|---|---|---|---|---|---|
| Instrument | Yes | Maybe | No | Yes | Maybe | No | Yes | Maybe | No |
| BLIS | 0.86 | 0.07 | 0.08 | 0.63 | 0.19 | 0.18 | 0.93 | 0.05 | 0.02 |
| M-BLIS | 0.84 | 0.07 | 0.09 | 0.58 | 0.21 | 0.21 | 0.94 | 0.03 | 0.03 |

Frequencies of responses are well-distributed across the two assessments on questions pertaining to the COVID-19 pandemic. The interest question (middle portion of the table) is the only one where some difference in proportion is observed for those selecting *Yes*, when separated by assessment type. However, the difference looks small enough.

Table 12 shows responses to the question "Did the context affect your ability to answer the statistical question?" This question will also be included in the models.

*Table 12. Self-reported effect of context on ability to respond to the statistical question*

| Context affecting ability | Percentage of respondents |
|---|---|
| No difference | 0.471 |
| Made it easier | 0.495 |
| Made it harder | 0.034 |

Finally, linear regression models were fit to explain the relationship between total scores and type of assessment. Additional variables were included to understand how the effect of assessment type on total score changed in the presence of other test-taker characteristics and their survey responses. This classical test theory-based approach towards analyzing assessment data was preferred due to its focus on the explanatory variables. Following a test for difference in means, various additive as well as interactive linear regression models incorporating a subset or all of the variables tabulated in Section 6.1 were fit. The adjusted $R^2$ for all models was in the proximity of 20%. We believe that for assessment data collected in an educational setting, this explanatory power is typical given the plethora of sources of variation. The final model discussed in this paper is per Equation ( 1 and the results presented in Table 21 (Section 6.4). This model includes all the covariates and had the highest adjusted $R^2$. For the results presented in Table 21, the base categories of the explanatory variables are marked with an asterix (*) in the univariate tables in Section 6.1.

$$\begin{aligned}
E(\text{Total score}) = \beta_0 &+ \beta_1 * \text{instrument} + \beta_2 * \text{international} + \beta_3 \ldots \beta_6 * \text{grade} + \beta_7 \\
&* \text{priorSTAT} + \beta_8 \ldots \beta_{10} * \text{class} + \beta_{11} \ldots \beta_{14} * \text{gender} + \beta_{15} \ldots \beta_{22} \\
&* \text{highest parent education} + \beta_{23} \ldots \beta_{24} * \text{COVID engagement} \\
&+ \beta_{25} \ldots \beta_{26} * \text{COVID interest} + \beta_{27} \ldots \beta_{28} * \text{COVID relevance} + \beta_{29} \\
&* \text{topic familiarity} + \beta_{30} * \text{topic interest} + \beta_{31} \ldots \beta_{32} \\
&* \text{effect of context on performance}
\end{aligned}$$

( 1 )

*Including all covariates and survey responses to the model increased the adjusted $R^2$ value to ~25%. We also considered a model with all the variables in Equation 1 interacting with instrument type. Almost all the interaction terms had high p-value with very little gain in the $R^2$ value (less than 1%), and therefore, they were not given further consideration. Results from this model are presented in* **Table 21: Results from the full regression model**

|  | Estimate | Std. Error | t-value | p-value |
|---|---|---|---|---|
| (Intercept) | 25.43 | 2.31 | 11.03 | 0.0000 |
| Instrument – M-BLIS | -1.21 | 0.33 | -3.64 | 0.0003 |
| International – Yes | -1.27 | 0.75 | -1.68 | 0.0937 |
| Grade – B | -4.57 | 0.38 | -11.95 | 0.0000 |
| Grade – C | -6.88 | 0.49 | -13.99 | 0.0000 |
| Grade – D | -8.77 | 0.99 | -8.87 | 0.0000 |
| Grade – F | -0.28 | 3.72 | -0.07 | 0.9407 |
| priorSTAT – Yes | 0.56 | 0.37 | 1.53 | 0.1255 |
| Class - Second Year (e.g.~Sophomore) | -0.31 | 0.41 | -0.76 | 0.4479 |
| Class - Third Year (e.g.~Junior) | -0.25 | 0.63 | -0.39 | 0.6980 |
| Class - Fourth Year or Higher (e.g.~Senior) | 1.77 | 1.02 | 1.74 | 0.0823 |
| Gender – Woman | -0.50 | 0.35 | -1.43 | 0.1524 |
| Gender – Transgender | -2.37 | 3.74 | -0.64 | 0.5256 |
| Gender – Prefer not to disclose | 2.84 | 3.67 | 0.77 | 0.4391 |
| Gender – Prefer to self-specify | 4.18 | 5.17 | 0.81 | 0.4182 |
| Highest parent ed – High school graduate | -2.72 | 1.96 | -1.39 | 0.1662 |
| Highest parent ed – College, no degree | -1.96 | 1.95 | -1.01 | 0.3144 |
| Highest parent ed – Associate's | -1.75 | 2.06 | -0.85 | 0.3966 |
| Highest parent ed – Bachelor's | -1.51 | 1.89 | -0.80 | 0.4236 |
| Highest parent ed – Some graduate school | 0.14 | 2.14 | 0.07 | 0.9470 |
| Highest parent ed – Master's | -0.70 | 1.90 | -0.37 | 0.7112 |
| Highest parent ed – Professional degree | -1.26 | 2.07 | -0.61 | 0.5415 |
| Highest parent ed – Doctorate | -0.41 | 1.95 | -0.21 | 0.8338 |
| COVID engagement – Maybe | -0.38 | 0.91 | -0.42 | 0.6726 |
| COVID engagement – Yes | 0.03 | 0.68 | 0.04 | 0.9681 |
| COVID interest – Maybe | 1.22 | 0.55 | 2.21 | 0.0274 |
| COVID interest – Yes | 0.39 | 0.49 | 0.80 | 0.4245 |
| COVID relevance – Maybe | -3.59 | 1.48 | -2.42 | 0.0155 |
| COVID relevance – Yes | -0.92 | 1.15 | -0.80 | 0.4260 |
| Familiarity with topic | -0.01 | 0.01 | -1.39 | 0.1655 |
| Interest in topic | 0.01 | 0.01 | 1.51 | 0.1324 |
| Context – made question easier | -0.17 | 0.35 | -0.48 | 0.6341 |
| Context – made question harder | -0.29 | 0.93 | -0.32 | 0.7520 |

Table 22 in the Appendix (Section 6.4).

No matter the model, instrument type was found to be related to the total score with a lower score on the modified assessment. For the model in Equation ( 1, the p-value for instrument point was 0.003. After accounting for the variation in scores explained by the instrument type, this models showed evidence of a relationship between some of the coviariates and survey responses, and total score. The strongest relationships, as indicated by small p-values, were with 1) grade expectations B, C, or D (base category 'A', lower scores, p-values 0.0000), 2) fourth year or higher students (base category 'First year students', higher scores, p-value 0.08), 3) 'Maybe' being interested in learning about COVID-19 through a statistical lens (base category 'Not' interested, higher score, p-value 0.03), and 4) 'Maybe' considering COVID-19 to be relevant to one's life (base category 'Not' relevant, lower score, p-value 0.02).

Diagnostic plots for these regression models are presented in the Appendix (Section 6.4). Since most of the explanatory variables we chose were categorical, scatter plots of the response variable or the residuals with the explanatory variables were not considered. The histogram of residuals looks fairly normal, with slight more density on the positive side. The scatterplot of fitted values versus residuals indicates a definitive pattern suggesting that there is an omitted variable bias in the results we are seeing. It may be reasonable to expect that additional variables at both test-taker and item level may be able to explain further variation. This may include racial and ethnic background, observed course performance, item text characteristics etc. Finally, the residual plots ascertained that heteroskedasticity is not a concern for these models.

## 4. DISCUSSION

This work demonstrates two important things. First, that a carefully designed isomorphic assessment can allow for reliable measurement of statistical literacy in specific contexts. Second, that a year into the COVID-19 pandemic (as of April 2021) students who were finishing-up a semester of college-level introductory statistics scored lower on a pandemic-specific assessment of statistical literacy as compared to another version with a variety of non-pandemic contexts. This lower score indicates that context matters!

These analyses inform two distinct questions at hand and what we learn from one informs the other. On the topic of isomorphic assessments, we set out to investigate whether the M-BLIS measures the same underlying constructs as BLIS and in the same way, or not. The reliability and validity analyses indicate that this is true for most items. Though, for the items where we notice a difference in factor loading or the item information curves, the differences are noticeable. Based on these, we can conclude that a carefully constructed isomorphic assessment can measure the same underlying constructs while exposing the test taker to statistical literacy concepts through the lens of a variety of application areas. For the items which indicate serious difference, our future work will look at factors such as reading difficulty as measured by a lexicon score that comprises of linguistic difficulty as well as length of text, whether the student is a native speaker of American english or not, whether they have prior statistics interest or not, and whether they are interested in studying the pandemic through a statistical lens or not. Responses will be analyzed to investigate which, if any, item or test-taker characteristics may be driving the differences in scores on the items with high differences in proportion of correct responses. The second question of interest to us was a comparison of student performance. These analyses were conducted assuming that scores on BLIS and M-BLIS are equatable under the internal-anchor design (Livingston, 2004). Various CTT-based analyses using multiple linear regression indicate that assessment type is an important predictor of total score no matter which other characteristics are included and whether the model includes any interactions or not.

### 4.1. LIMITATIONS

In general, the less portable items on BLIS required either raw quantitative data, data from a randomized experiment, or data that led to visualizations with peculiar characteristics such as strong right skewness. Concessions were made in case of three items where for one item the parameter of interest was switched from mean to proportion, and an observational study was discussed instead of a randomized experiment in one other. As seen in the example in Table 2, reverse skewness was accepted for a third item. However, the lack of open availability of raw datasets is a hurdle that will need to be addressed more systematically in creating future isomorphs.

Additionally, balancing the competing goals of maximizing engagement and minimizing emotional impact lead to the inclusion of some topics which may not be most relevant to the lives of of our target population for the study - college students, in this case - and exclusion of some topics which may be directly related to them. For example, one of the modified items referred to pre- and during pandemic performance of elementary school students on standardized tests. This issue is confounded by the expectations of the 'college student' audience which is typical to an educational research study, though that may not need to be the case for the general purpose of the research. The choice of the test population can bias the choice of relevant contexts.

The item in Table 5 was a subject of lengthy discussions, some of which included the expert reviewers. The implicit assumption of a coin being unbiased and our intuition about 50% of them landing on heads benefitted the original item. However, upon deliberation, it was agreed that it is extremely hard to find other phenomena which have an unconditional 0.5 probability of occurrence which is understood intuitively, and therefore the substantial change in wording was included. The original item was an interesting case because students are assumed to be so familiar with fair coins that the frequency of their 'encounters' with the context might actually outweigh the other dimensions of engagement/relevance we are seeking in this study.

Authors must also acknowledge that even though we use anchor items to compare the two sets of respondents at baseline, we have to account for possible ordering effect. These identical items could function differently across BLIS and M-BLIS, especially since they may appear out-of-context on an assessment based entirely on one specific topic – the COVID-19 pandemic.

Finally, survey questions were asked at the end of the assessment. Therefore, we didn't expect that students' performance on the assessment would have been affected by these.

However, responses to the survey questions may have contained some cognitive bias based on whether they had just seen an entire assessment based on COVID-19 or not.

### 4.2. IMPLICATION FOR FUTURE WORK

Since the instruments are observed to function comparably, we argue that isomorphic assessment can be created to assess statistical literacy in various pertinent contexts. Even though it may be quite tedious create them, these instruments can be invaluable tools in getting respondents to consider statistics through a contextual lens that is relevant, and continue to measure how curricular strategies may affect literacy levels. Therefore, future research can be directed towards two purposes. 1) measurement of statistical literacy in various disciplinary or societal contexts using isomorphs of BLIS, and 2) using these isomorphic versions to assess performance of experimental curricular or pedagogical strategies. However, additional work exploring the transfer and cognitive processes behind statistical problem solving will also be essential to our understanding the role of contexts.

The pilot study was intended to study psychometric properties of M-BLIS in comparison with BLIS to determine whether the BLIS and M-BLIS are psychometrically isomorphic, and whether they measure the same constructs even when the context is changed. To draw reliable conclusions, it was essential that we have the ability to compare results from our study to the field test conducted during the development of the original assessment. To achieve this, it was important to ensure that the BLIS items remained identical to that test, and therefore, M-BLIS was based on that version. At no point did we change any details in the original assessment in an effort to ensure comparability across the original work (Ziegler, 2014) and our pilot study. Resultantly, the results from this paper are specific to one definition and assessment of statistical literacy. Future research should study the role of contexts using other asssessment instruments.

Differential student performance on BLIS and M-BLIS with a low p-value on inferential results indicates that the context in which a statistical question is posed affects assessment responses. Our respondents made sense of statistical questions differently based on whether the context behind the numbers was relevant to them or not. In reference to the discussion in Section 2.1 regarding sensitive contexts, this finding also has implications for teaching practices. If additional research finds that the sensitivity of the topic may have contributed to the lower scores on M-BLIS, an argument can be made to favor inclusion of such topics on curricular materials instead of including them in grade-affecting assessments (Fallstrom et al., 2021).

From a context point of view, two things are worth noting. First, as discussed in Section 3.1, some of the BLIS items pertaining to college students saw better performance even though the examples were realistic. This may suggest that relevance itself may be hypercontextualized for different subgroups. Secondly, it was interesting to note in Table 11 that there was a certain percentage of students who, no matter which assessment they took, indicated after completing the assessment that they had engaged with the COVID-19 pandemic by seeking out information, believed it was relevant to their lives, yet would not be interested in gaining data-driven insights into the pandemic. Granted, this study ran about 13 months into the pandemic and there may have been pandemic fatigue. However, this was at the end of a semester during which they had taken an introductory statistics class, making this an interesting phenomena warranting further investigation.

## 5. CONCLUSION

For a statistically literate individual, the ability to marry one's understanding of statistical constructs and the context-at-hand is assumed. In fact, as there is no statistics without context (Rao, 1975), statistical literacy is also inherently contextualized. However, the transfer of statistical skills to new contexts is non-trivial and this work further examines how contexts may factor into test-takers' responses to a research-based assessment of statistical literacy. Parallel to Greer et al. (2007)'s discussion in the context of mathematics education, statistics education, too, is a way for us to develop citizens who can make sense of quantitative information in contexts that matter to them. Towards that goal, this work will allow researchers to better understand how students of statistics showcase statistical literacy skills in the context of relevant topics, and inform instructional practices which can maximize such transfer in the future.

## ACKNOWLEDGEMENTS

Work supported, in part, by Penn State's Centre for Social Data Analytics Accelerator Award Program Seed Grant. Any opinions, findings, and conclusions or recommendations are those of the authors and do not necessarily reflect those of the sponsor. We wish to acknowledge the three expert reviewers of M-BLIS who provided valuable input. We wish to also thank Dr. Ziegler for her support during the adaptation of her original work.

SAYALI PHADKE
445 Waupelani Drive,
Apartment J25,
State College,
PA 16801
USA

# 6. APPENDIX

## 6.1. APPENDIX A: RESPONDENT DEMOGRAPHICS

*Univariate summaries*

*Table 13. Gender identification: n = 1253*

| Gender | Frequency |
|---|---|
| Woman | 0.605 |
| Man* | 0.387 |
| Transgender | 0.003 |
| Prefer not to disclose | 0.003 |
| Prefer to self-specify | 0.002 |

*Table 14. International student: n = 1253*

| Whether an international student | Frequency |
|---|---|
| No* | 0.931 |
| Yes | 0.069 |

*Table 15. Class standing: n = 1253*

| Class standing | Frequency |
|---|---|
| First Year (e.g.~Freshman)* | 0.638 |
| Second Year (e.g.~Sophomore) | 0.228 |
| Third Year (e.g.~Junior) | 0.097 |
| Fourth Year or Higher (e.g.~Senior) | 0.037 |

*Table 16. Prior statistics training: n = 1253*

| Prior statistics training | Frequency |
|---|---|
| No* | 0.696 |
| Yes | 0.304 |

*Table 17. Expected course grade: n = 1253*

| Expected course grade | Frequency |
|---|---|
| A* | 0.338 |
| B | 0.410 |
| C | 0.216 |
| D | 0.032 |
| F | 0.004 |

*Table 18. Highest education of parent/guardian: n = 1253*

| Highest education level of a parent/guardian | Frequency |
|---|---|
| Less than high school* | 0.011 |
| High school graduate | 0.080 |
| Some college, no degree | 0.080 |
| Associates Degree | 0.041 |
| Bachelor's Degree | 0.391 |
| Some graduate school | 0.025 |
| Master's Degree | 0.260 |
| Professional Degree | 0.036 |
| Doctorate Degree | 0.076 |

## 6.2. ASSESSMENT RESPONSE SUMMARIES

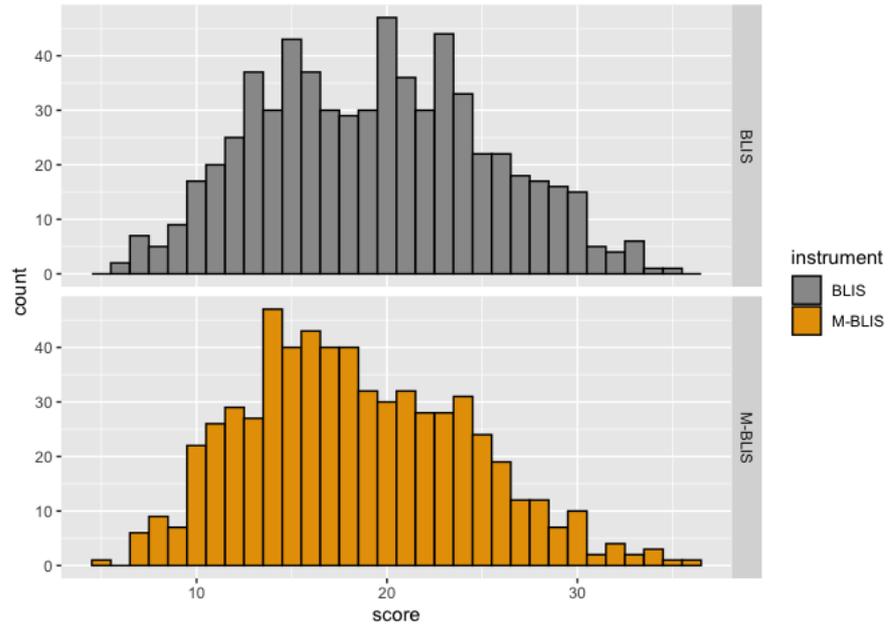

*Figure 3: Comparison of total score (out of 37) - separate panels*

*Table 19. Selected-response table*

| | BLIS | M-BLIS |
|---|---|---|

| Item | A | B | C | D | A | B | C | D |
|---|---|---|---|---|---|---|---|---|
| 1 | 16.6 | 8.8 | 74.6* | NA | 19.3 | 7.5 | 73.2* | NA |
| 2 | 7.4 | 15.8 | 44* | 32.8 | 8.6 | 18 | 50.7* | 22.6 |
| 3 | 53.4* | 20.5 | 7.2 | 18.8 | 52.5* | 8.6 | 13 | 25.9 |
| 4 | 15.7 | 83.5* | 0.8 | NA | 13 | 86.2* | 0.8 | NA |
| 5 | 81.3* | 16.5 | 2.2 | NA | 84.7* | 13.8 | 1.5 | NA |
| 6 | 3.4 | 10.8 | 73.5* | 12.2 | 4.7 | 15.8 | 70.7* | 8.8 |
| 7 | 32.6 | 15.7 | 16.1 | 35.6* | 21 | 17.1 | 20.8 | 41.1* |
| 8 | 39.8 | 30.7 | 29.5* | NA | 23.7 | 43.4 | 32.8* | NA |
| 9 | 65.4* | 30.7 | 3.9 | NA | 34* | 56.4 | 9.6 | NA |
| 10 | 14.7 | 9.1 | 19.9 | 56.3* | 21.6 | 21.5 | 17.7 | 39.2* |
| 11 | 19.1 | 38.9 | 42* | NA | 18.5 | 44.4 | 37.1* | NA |
| 12 | 13.9 | 22.7 | 5 | 58.3* | 11.9 | 31.1 | 8.3 | 48.8* |
| 13* | 37.6* | 40.3 | 8.5 | 13.6 | 37.6* | 35 | 14.1 | 13.3 |
| 14 | 5.8 | 42.8* | 50.3 | 1.1 | 2.8 | 69.1 | 24.6* | 3.6 |
| 15 | 5.6 | 8.9 | 21.6 | 63.8* | 6 | 18.2 | 27.3 | 48.5* |
| 16* | 24.6* | 19.9 | 36.1 | 19.4 | 27.8* | 42.4 | 29.8 | NA |
| 17* | 25.2 | 28.7 | 46.1* | NA | 23.7 | 29.4 | 46.8* | NA |
| 18 | 35.3 | 45.9* | 18.8 | NA | 29.9 | 45.4* | 24.7 | NA |
| 19 | 9.7 | 29.8 | 19.7 | 40.8* | 11.1 | 27.6 | 22.9 | 38.4* |
| 20 | 20.7 | 29 | 37.9* | 12.4 | 17.9 | 34.8 | 34.3* | 13 |
| 21 | 16.5* | 8.8 | 39.3 | 35.4 | 16.3* | 14.1 | 36.7 | 32.8 |
| 22 | 29.6 | 58.5* | 11.9 | NA | 26.8 | 61* | 12.2 | NA |
| 23* | 12.2 | 29.8 | 43.4* | 14.6 | 10.4 | 31.1 | 43.9* | 14.6 |
| 24* | 57.2* | 25.4 | 17.4 | NA | 60* | 22.4 | 17.6 | NA |
| 25 | 55.5* | 25.9 | 18.7 | NA | 61.6* | 28 | 10.4 | NA |
| 26 | 42.2* | 42.5 | 15.4 | NA | 42* | 42.3 | 15.8 | NA |
| 27 | 31.7 | 38.6* | 18.8 | 11 | 27.5 | 45* | 14.6 | 12.8 |
| 28 | 20.5 | 26.8 | 52.7* | NA | 17.2 | 22.6 | 60.2* | NA |
| 29 | 18 | 17.4 | 52* | 12.5 | 13.3 | 20.2 | 48.9* | 17.6 |
| 30 | 10 | 19.9 | 21.8 | 48.3* | 9.6 | 22.4 | 22.4 | 45.5* |
| 31 | 8.2 | 86.4* | 5.5 | NA | 9.9 | 83.9* | 6.2 | NA |
| 32* | 19.7 | 20.8 | 48* | 11.4 | 20 | 23.6 | 43.6* | 12.8 |
| 33 | 64.4* | 22.1 | 13.5 | NA | 62* | 25.5 | 12.5 | NA |
| 34 | 25.4 | 70.4* | 4.2 | NA | 31.2 | 65.2* | 3.6 | NA |
| 35 | 23.4* | 11.4 | 13 | 52.2 | 21.6* | 16.9 | 15.4 | 46 |
| 36 | 79* | 16.6 | 4.4 | NA | 68.6* | 23.7 | 7.6 | NA |
| 37 | 16.6 | 20.7 | 57.8* | 4.9 | 19.8 | 20.8 | 54.3* | 5 |

## 6.3. RELIABILITY AND VALIDITY EVIDENCE

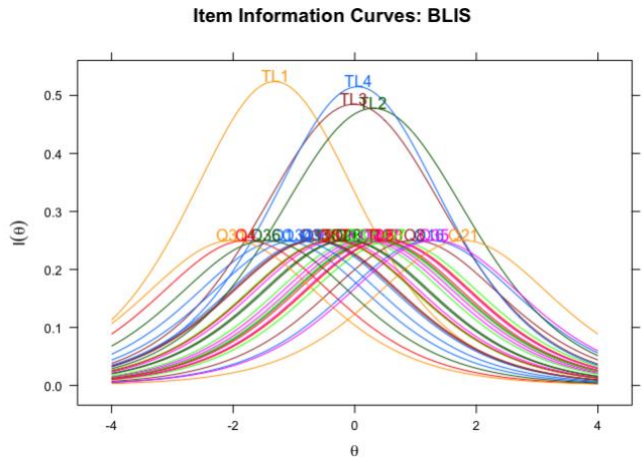

*Figure 4. Item Information Curves - BLIS*

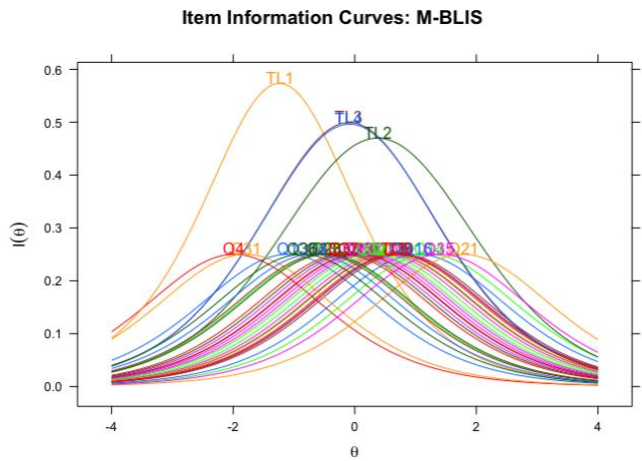

*Figure 5. Item Information Curves - M-BLIS*

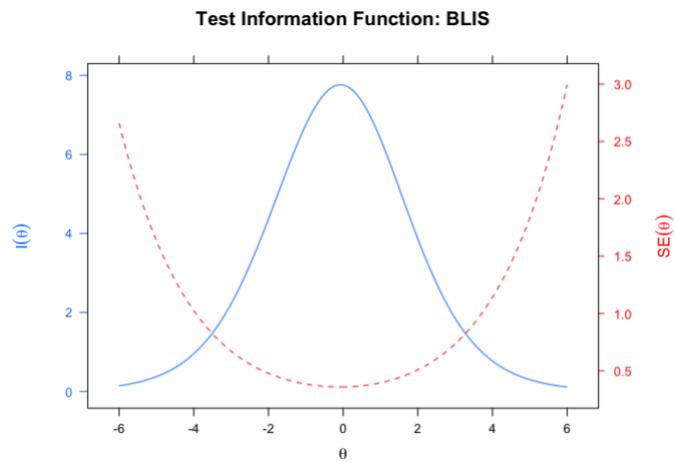

*Figure 6. Test Information Function and Standard Error - BLIS*

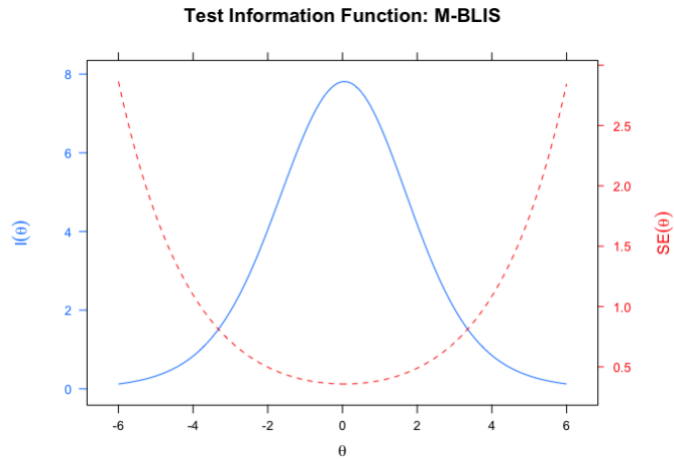

*Figure 7. Test Information Function and SE - M-BLIS*

*Table 20: Difficulty estimates based on PC model*

| Item | BLIS | M-BLIS |
|------|------|--------|
| Q1 | -1.1927156 | -1.1021233 |
| Q2 | 0.2631418 | -0.0375388 |
| Q3 | -0.1571817 | -0.1164450 |
| Q4 | -1.7811054 | -1.9824458 |
| TL1 | -1.3170586 | -1.2386935 |
| Q7 | 0.6567262 | 0.3897469 |
| Q8 | 0.9650827 | 0.7822983 |
| Q9 | -0.7077156 | 0.7261573 |
| Q10 | -0.2838843 | 0.4792098 |
| Q11 | 0.3556544 | 0.5778552 |
| Q12 | -0.3763797 | 0.0485164 |
| Q13 | 0.5593310 | 0.5549103 |
| Q14 | 0.3199521 | 1.2271370 |
| Q15 | -0.6321048 | 0.0628702 |
| Q16 | 1.2357476 | 1.0437791 |
| Q17 | 0.1714903 | 0.1347889 |
| TL2 | 0.3055317 | 0.3746852 |
| Q20 | 0.5445291 | 0.7102782 |
| Q21 | 1.7825288 | 1.7836447 |
| Q22 | -0.3835169 | -0.4960496 |
| TL3 | -0.0174983 | -0.0913839 |
| TL4 | 0.0476351 | -0.0845298 |
| Q27 | 0.5150569 | 0.2142002 |
| Q28 | -0.1221481 | -0.4587553 |
| TL5 | 0.4059238 | 0.6708731 |
| Q31 | -2.0158467 | -1.7937093 |
| Q32 | 0.0873939 | 0.2796047 |
| Q33 | -0.6621846 | -0.5411665 |
| Q34 | -0.9608257 | -0.6947486 |
| Q35 | 1.3108151 | 1.4059160 |
| Q36 | -1.4599263 | -0.8628098 |

| Q37 | -0.3549298 | -0.1956216 |

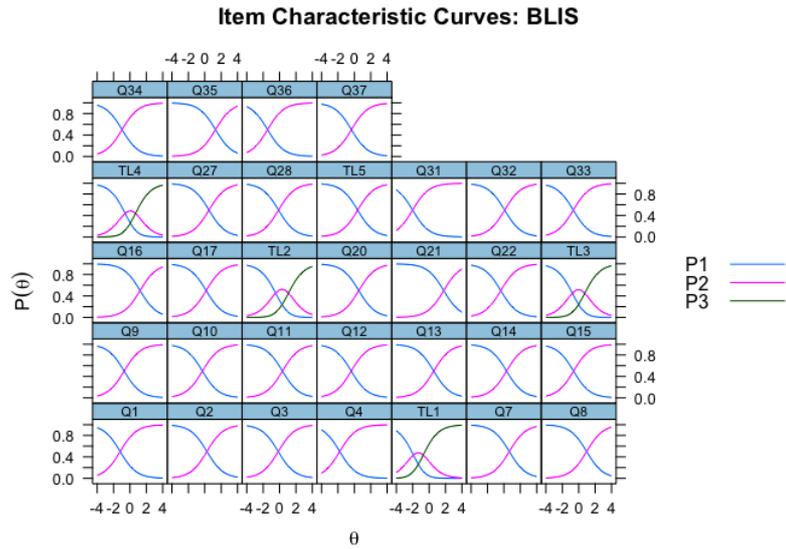

*Figure 8. Item Characteristic Curves - BLIS*

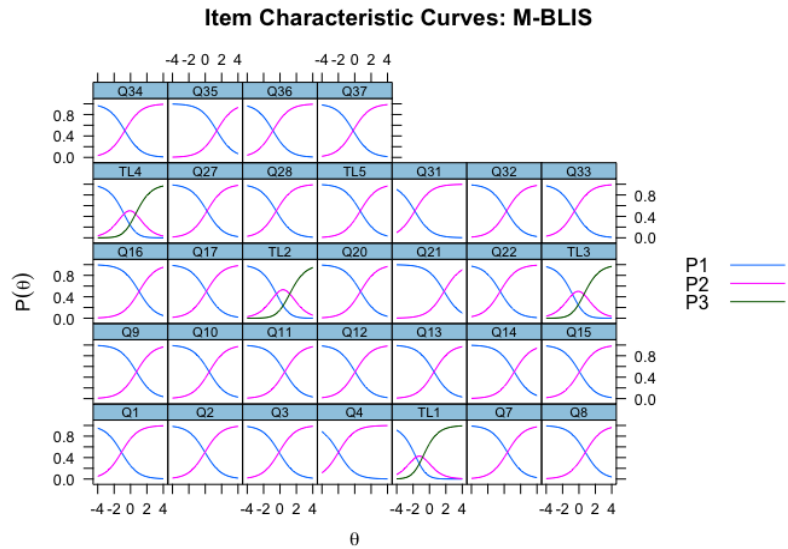

*Figure 9. Item Characteristic Curves - M-BLIS*

## 6.4. REGRESSION RESULTS

*Table 21: Results from the full regression model*

|  | Estimate | Std. Error | t-value | p-value |
|---|---|---|---|---|
| (Intercept) | 25.43 | 2.31 | 11.03 | 0.0000 |
| Instrument – M-BLIS | -1.21 | 0.33 | -3.64 | 0.0003 |
| International – Yes | -1.27 | 0.75 | -1.68 | 0.0937 |
| Grade – B | -4.57 | 0.38 | -11.95 | 0.0000 |
| Grade – C | -6.88 | 0.49 | -13.99 | 0.0000 |
| Grade – D | -8.77 | 0.99 | -8.87 | 0.0000 |
| Grade – F | -0.28 | 3.72 | -0.07 | 0.9407 |
| priorSTAT – Yes | 0.56 | 0.37 | 1.53 | 0.1255 |
| Class - Second Year (e.g.~Sophomore) | -0.31 | 0.41 | -0.76 | 0.4479 |
| Class - Third Year (e.g.~Junior) | -0.25 | 0.63 | -0.39 | 0.6980 |
| Class - Fourth Year or Higher (e.g.~Senior) | 1.77 | 1.02 | 1.74 | 0.0823 |
| Gender – Woman | -0.50 | 0.35 | -1.43 | 0.1524 |
| Gender – Transgender | -2.37 | 3.74 | -0.64 | 0.5256 |
| Gender – Prefer not to disclose | 2.84 | 3.67 | 0.77 | 0.4391 |
| Gender – Prefer to self-specify | 4.18 | 5.17 | 0.81 | 0.4182 |
| Highest parent ed – High school graduate | -2.72 | 1.96 | -1.39 | 0.1662 |
| Highest parent ed – College, no degree | -1.96 | 1.95 | -1.01 | 0.3144 |
| Highest parent ed – Associate's | -1.75 | 2.06 | -0.85 | 0.3966 |
| Highest parent ed – Bachelor's | -1.51 | 1.89 | -0.80 | 0.4236 |
| Highest parent ed – Some graduate school | 0.14 | 2.14 | 0.07 | 0.9470 |
| Highest parent ed – Master's | -0.70 | 1.90 | -0.37 | 0.7112 |
| Highest parent ed – Professional degree | -1.26 | 2.07 | -0.61 | 0.5415 |
| Highest parent ed – Doctorate | -0.41 | 1.95 | -0.21 | 0.8338 |
| COVID engagement – Maybe | -0.38 | 0.91 | -0.42 | 0.6726 |
| COVID engagement – Yes | 0.03 | 0.68 | 0.04 | 0.9681 |
| COVID interest – Maybe | 1.22 | 0.55 | 2.21 | 0.0274 |
| COVID interest – Yes | 0.39 | 0.49 | 0.80 | 0.4245 |
| COVID relevance – Maybe | -3.59 | 1.48 | -2.42 | 0.0155 |
| COVID relevance – Yes | -0.92 | 1.15 | -0.80 | 0.4260 |
| Familiarity with topic | -0.01 | 0.01 | -1.39 | 0.1655 |
| Interest in topic | 0.01 | 0.01 | 1.51 | 0.1324 |
| Context – made question easier | -0.17 | 0.35 | -0.48 | 0.6341 |
| Context – made question harder | -0.29 | 0.93 | -0.32 | 0.7520 |

*Table 22. Results from the full regression model with interactions*

| | Estimate | Std. Error | t-value | p-value |
|---|---|---|---|---|
| (Intercept) | 28.89 | 3.83 | 7.53 | 0.0000 |
| Instrument – M-BLIS | -5.80 | 4.81 | -1.20 | 0.2286 |
| International – Yes | 0.0788 | 1.1140 | 0.0708 | 0.8752 |
| Grade – B | -4.5921 | 0.5257 | -8.7348 | 0.0000 |
| Grade – C | -6.7615 | 0.6682 | -10.1184 | 0.0000 |
| Grade – D | -8.5514 | 1.2567 | -6.8048 | 0.0000 |
| Grade – F | 0.4755 | 3.7834 | 0.1257 | 0.9096 |
| priorSTAT – Yes | 1.1996 | 0.4903 | 2.4465 | 0.0172 |
| Class - Second Year (e.g.~Sophomore) | 0.2548 | 0.5755 | 0.4428 | 0.8956 |
| Class - Third Year (e.g.~Junior) | -0.2159 | 0.8732 | -0.2472 | 0.6428 |
| Class - Fourth Year or Higher (e.g.~Senior) | 1.2674 | 1.4629 | 0.8664 | 0.0870 |
| Gender – Woman | -0.2991 | 0.4722 | -0.6335 | 0.4736 |
| Gender – Transgender | -1.4369 | 3.8185 | -0.3763 | 0.7167 |
| Gender – Prefer not to disclose | 3.4160 | 3.6900 | 0.9257 | 0.3400 |
| Gender – Prefer to self-specify | 4.0562 | 5.1820 | 0.7827 | 0.4726 |
| Highest parent ed – High school graduate | -3.2444 | 2.8007 | -1.1584 | 0.0548 |
| Highest parent ed – College, no degree | -3.8873 | 2.8068 | -1.3850 | 0.0329 |
| Highest parent ed – Associate's | -4.0512 | 2.9570 | -1.3700 | 0.0376 |
| Highest parent ed – Bachelor's | -3.0528 | 2.7370 | -1.1154 | 0.0584 |
| Highest parent ed – Some graduate school | 0.3851 | 3.0209 | 0.1275 | 0.5557 |
| Highest parent ed – Master's | -1.4848 | 2.7361 | -0.5427 | 0.1755 |
| Highest parent ed – Professional degree | -0.9369 | 2.9184 | -0.3210 | 0.2416 |
| Highest parent ed – Doctorate | -0.9873 | 2.8362 | -0.3481 | 0.2580 |
| COVID engagement – Maybe | 0.1007 | 1.2342 | 0.0816 | 0.7519 |
| COVID engagement – Yes | 0.6973 | 0.9523 | 0.7322 | 0.2651 |
| COVID interest – Maybe | 0.6063 | 0.7670 | 0.7904 | 0.4399 |
| COVID interest – Yes | 0.2437 | 0.6822 | 0.3573 | 0.8167 |
| COVID relevance – Maybe | -4.5675 | 2.1554 | -2.1191 | 0.0688 |
| COVID relevance – Yes | -2.6750 | 1.8918 | -1.4140 | 0.1646 |
| Familiarity with topic | 0.0000 | 0.0136 | -0.0010 | 0.8783 |
| Interest in topic | 0.0100 | 0.0129 | 0.7748 | 0.5249 |
| Context – made question easier | 0.7601 | 0.4796 | 1.5850 | 0.1034 |
| Context – made question harder | -0.2331 | 1.3634 | -0.1710 | 0.8917 |
| M-BLIS * International – Yes | -2.1982 | 1.5346 | -1.4324 | 0.1621 |
| M-BLIS * Grade – B | 0.1978 | 0.7571 | 0.2613 | 0.9379 |
| M-BLIS * Grade – C | 0.0312 | 0.9662 | 0.0323 | 0.9841 |
| M-BLIS * Grade – D | -0.4991 | 1.9815 | -0.2519 | 0.7459 |
| M-BLIS * priorSTAT – Yes | -1.4494 | 0.7287 | -1.9891 | 0.0687 |
| M-BLIS * Class - Second Year | -1.1473 | 0.8154 | -1.4071 | 0.3645 |
| M-BLIS * Class - Third Year | -0.2476 | 1.2452 | -0.1989 | 0.8984 |
| M-BLIS * Class - Fourth Year or Higher | 0.0364 | 1.9530 | 0.0186 | 0.4287 |
| M-BLIS * Gender – Woman | -0.3288 | 0.6920 | -0.4751 | 0.5639 |
| M-BLIS * Parent ed – High school graduate | 1.4023 | 3.8061 | 0.3684 | 0.2728 |
| M-BLIS * Parent ed – College, no degree | 4.1724 | 3.7799 | 1.1038 | 0.0754 |
| M-BLIS * Parent ed – Associate's | 5.3618 | 4.0086 | 1.3376 | 0.0519 |
| M-BLIS * Parent ed – Bachelor's | 3.2610 | 3.6685 | 0.8889 | 0.1024 |
| M-BLIS * Parent ed – Some graduate school | -1.3695 | 4.1521 | -0.3298 | 0.7980 |
| M-BLIS * Parent ed – Master's | 1.8308 | 3.6731 | 0.4984 | 0.2220 |
| M-BLIS * Parent ed – Professional degree | 0.2210 | 4.0395 | 0.0547 | 0.4898 |
| M-BLIS * Parent ed – Doctorate | 1.6842 | 3.7887 | 0.4445 | 0.2900 |
| M-BLIS * COVID engagement – Maybe | -1.4961 | 1.7527 | -0.8536 | 0.4441 |
| M-BLIS * COVID engagement – Yes | -1.2941 | 1.3267 | -0.9755 | 0.1871 |
| M-BLIS * COVID interest – Maybe | 0.9790 | 1.0848 | 0.9025 | 0.2342 |
| M-BLIS * COVID interest – Yes | -0.0937 | 0.9731 | -0.0963 | 0.7556 |
| M-BLIS * COVID relevance – Maybe | 0.8075 | 3.0687 | 0.2631 | 0.9063 |
| M-BLIS * COVID relevance – Yes | 2.9060 | 2.3986 | 1.2115 | 0.2139 |
| M-BLIS * Familiarity with topic | -0.0179 | 0.0201 | -0.8892 | 0.4097 |
| M-BLIS * Interest in topic | 0.0073 | 0.0190 | 0.3855 | 0.7339 |
| M-BLIS * Context – made question easier | -1.9502 | 0.6933 | -2.8130 | 0.0065 |
| M-BLIS * Context – made question harder | -0.5543 | 1.8714 | -0.2962 | 0.5633 |

***Diagnostic plots for the additive model***

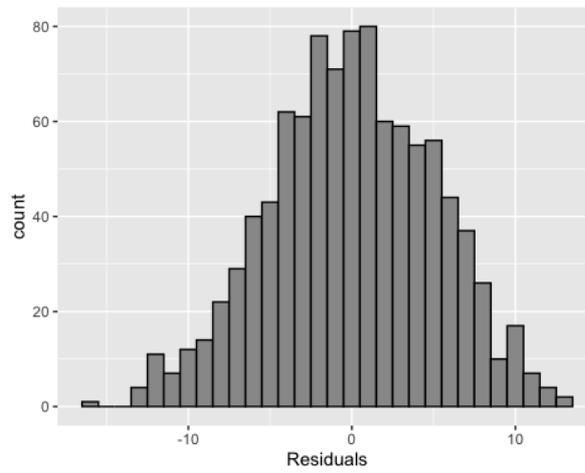
*Figure 10. Histogram of residuals - Full model in Equation 1*

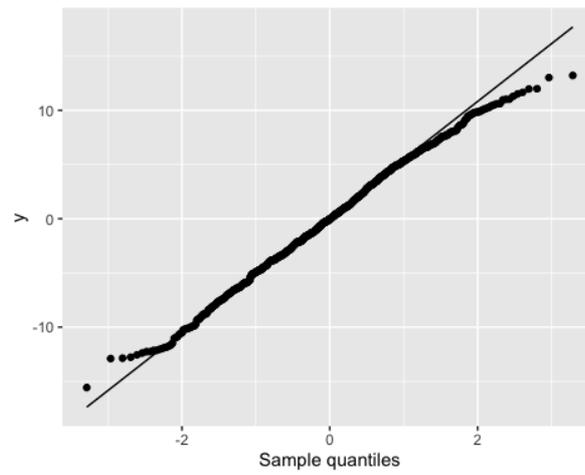
*Figure 11. Quartile-quartile plot of residuals - Full model in Equation 1*

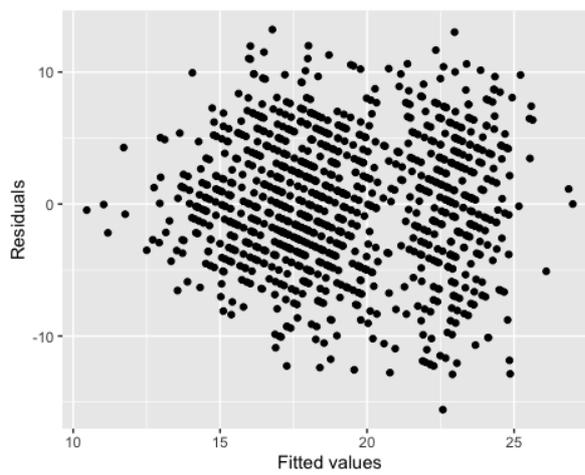
*Figure 12. Fitted values versus residuals plot - Full model in Equation 1*

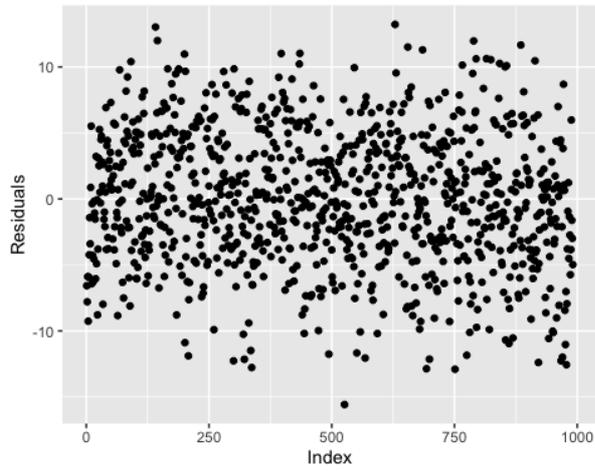

*Figure 13. Residuals plot - Full model in Equation 1*

***Diagnostic plots for the model with interactions***

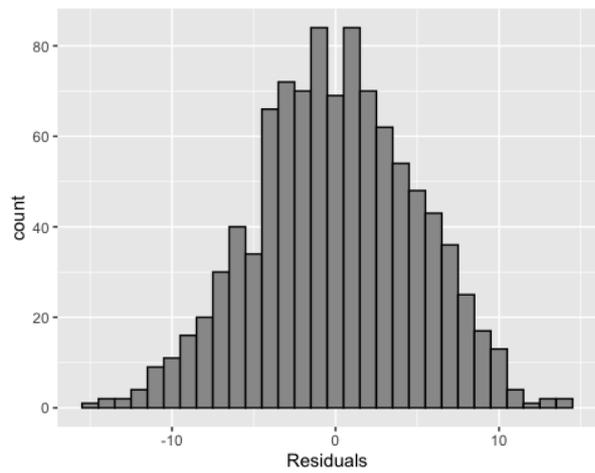

*Figure 14. Histogram of residuals - Full model plus interactions*

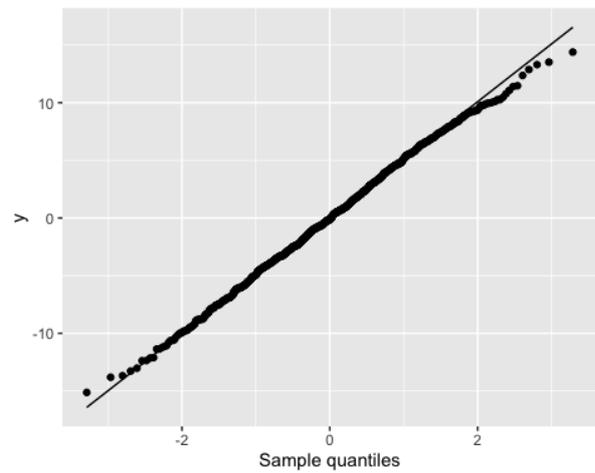

*Figure 15. Quartile-quartile plot of residuals - Full model plus interactions*

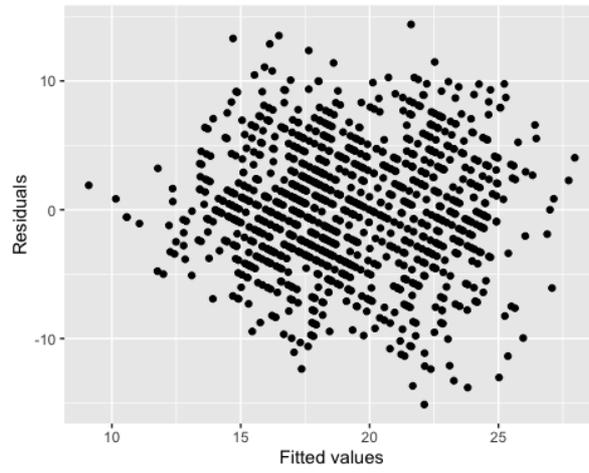

*Figure 16. Fitted values versus residuals plot - Full model plus interactions*

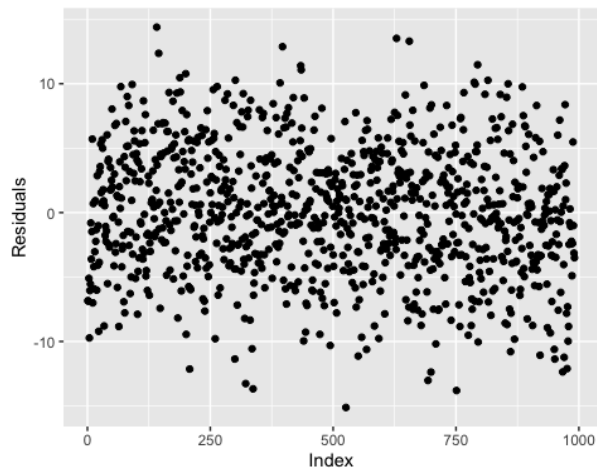

*Figure 17. Residuals plot - Full model plus interactions*